\documentclass[twocolumn,showpacs,amsmath,amssymb,prl]{revtex4}

\usepackage{graphicx}
\usepackage{dcolumn}
\usepackage{bm}

\newcommand{\adj}{+}
\newcommand{\am}{a\,}
\renewcommand{\ap}{a^{\adj}} 
\newcommand{\B}{\boldsymbol{B}}
\newcommand{\Bx}{B_{x}}
\newcommand{\By}{B_{y}}
\newcommand{\Bz}{B_{z}}
\newcommand{\Bcp}{B_{+}}
\newcommand{\Bcm}{B_{-}}
\newcommand{\Bcpm}{B_{\pm}}

\newcommand{\bi}{\boldsymbol{q}}
\newcommand{\D}{{\rm d}}
\newcommand{\dm}{\varrho}
\newcommand{\e}{\mathrm{e}}
\newcommand{\el}{\varepsilon}
\newcommand{\half}{\tfrac{1}{2}}
\newcommand{\Ham}{{\cal H}}
\newcommand{\Hs}{{\cal H}}
\newcommand{\iu}{\mathrm{i}}
\newcommand{\Ker}{{\cal K}}
\newcommand{\kT}{T}
\newcommand{\lf}{\ell}
\newcommand{\LF}{L}
   
\newcommand{\llangle}{\left\langle}
\newcommand{\n}{n}
\newcommand{\nbra}{\langle\n|}
\newcommand{\nket}{|\n\rangle}
\newcommand{\m}{m}
\newcommand{\mbra}{\langle\m|}
\newcommand{\mket}{|\m\rangle}
\newcommand{\mc}{\boldsymbol{c}}

\newcommand{\mQ}{\mathbb{\Q}}

\newcommand{\Q}{Q}
\newcommand{\rrangle}{\right\rangle}
\newcommand{\s}{\boldsymbol{S}}
\newcommand{\Scp}{S_{+}}
\newcommand{\Scm}{S_{-}}
\newcommand{\Scpm}{S_{\pm}}
\newcommand{\Si}{S_{i}}
\newcommand{\Sj}{S_{j}}
\newcommand{\Sk}{S_{k}}
\newcommand{\Sx}{S_{x}}
\newcommand{\Sy}{S_{y}}
\newcommand{\Sz}{S_{z}}
\newcommand{\tf}{\Delta}
\newcommand{\tm}{|}
\newcommand{\tp}{\tau}
\newcommand{\Tr}{\mathrm{Tr}}
\newcommand{\Vz}{v}
\newcommand{\w}{\Omega}
\newcommand{\wa}{\Omega_{\mathrm{a}}}
\newcommand{\Wu}{W}      
\newcommand{\Wo}{\lambda}
\newcommand{\X}{\boldsymbol{X}}
\newcommand{\Z}{{\cal Z}}

\begin{document}

\bibliographystyle{prsty}


\title{
Spin dynamics in a  dissipative environment: from quantal to classical
}

\author{
J. L. Garc\'{\i}a-Palacios
and
S. Dattagupta
}
\affiliation{%
S.~N.~Bose National Centre for Basic Sciences,
Salt Lake, Calcutta - 700\,098, India
}%

\date{\today}

\begin{abstract}
We address the problem of spin dynamics in the presence of a thermal
bath, by solving exactly the appropriate quantum master equations with
continued-fraction methods.
The crossover region between the quantum and classical domains is
studied by increasing the spin value $S$, and the asymptote for the
classical absorption spectra is eventually recovered.
Along with the recognized relevance of the coupling strength, we show
the critical role played by the {\em structure\/} of the
system-environment interaction in the emergence of classical
phenomenology.
\end{abstract}


\maketitle



Quantum mechanics is one of the most subtle and powerful theoretical
constructions of the human mind.
Understanding its implications, relation with other theories, and
domain of validity has captivated scientists since its advent.
This domain has been slowly expanding from the traditional of atoms and
molecules, to condensed matter systems (solids and liquids), and more
recently by studies of decoherence, quantum analogs of classical
effects (e.g., chaos), and the quantum-to-classical transition
\cite{zur91}.

These studies have also brought an increasing awareness of the role of
the environment.
Thus, the field of {\em open quantum systems\/} deals with systems
consisting of a few relevant degrees of freedom coupled to the
surrounding medium, which has a large number of constituents (photons,
phonons, electrons, nuclei, etc.).
The coupling produces dissipation, fluctuations, and decoherence; it
also enables the system to interchange energy and correlations with
the bath and relax to equilibrium \cite{weiss}.
Besides its basic interest, the above generic conditions make this
topic relevant in various areas of physics and chemistry.

Spins constitute one of the most paradigmatic quantum systems due to
their discrete and finite energy spectrum.
Their dynamics is also special and rich because of the underlying
commutation relations
$[\Si\,,\Sj]=\iu\,\epsilon_{ijk}\Sk$.
Naturally, it is important to take into account environmental effects
in spin problems, and this has led to several theories of {\em spin
relaxation}.
To deal rigorously with quantum dissipative systems, however, is a
difficult task.
Path integral propagators, quantum Langevin or master equations can
typically be solved in a few simple cases: free particle (or in a
uniform field), harmonic oscillator \cite{weiss}, two-state systems
(e.g., $S=1/2$ spins) \cite{legetal87}, etc.

The {\em continued-fraction method}, devised originally for classical
Brownian motion \cite{risken}, has been successfully adapted to solve
master equations for some quantum systems
\cite{allaribam77,shiuch93,vogris88,gar2004}.
Here we shall apply this technique to a spin with {\em arbitrary\/} $S$
weakly coupled to a dissipative bath, and monitor its intrinsic
dynamics via spin resonance.
We investigate how the approach to the classical results takes place
(out of reach of previous exact methods due to their limitations in
$S$).
We focus on the effects of the environment, not only of the coupling
strength, but also of the {\em structure\/} of the spin-bath Hamiltonian.
Usual studies of open quantum systems overlook the latter and adopt the
simplest bilinear interaction.
We consider two models with a solid-state motivation: coupling to
electron-hole excitations, actually linear in $\s$, and to phonons, an
even polynomial in $\s$.
We find that the approach to the classical results depends {\em
qualitatively\/} on the coupling structure.
This is specially critical for the uniformity of the convergence in
the different frequency sectors of the spin absorption spectra.
The problem is not merely academic; large-spin molecular clusters are
in the focus, while magnetic nanoparticles provide a natural classical
limit \cite{blupra2004}.
Thus, our results could also help discriminating different proposed
couplings in those systems and hence ascertain the microscopic origin
of dissipation.


Let us start with the Hamiltonian of a spin $\s$ coupled to a
bosonic bath (linearly in the bath variables)
%
\begin{equation}
\label{H}
\Ham_{\rm tot}
=
\Hs(\s)
+
{\textstyle\sum}_{\bi}
\,
V_{\bi}
F_{\bi}(\s)
\left(\ap_{\bi}+\am_{-\bi}\right)
+
\Ham_{\rm b}
\;.
\end{equation}
Here $\Hs(\s)$ and
$\Ham_{\rm b}=\sum_{\bi}\omega_{\bi}\,\ap_{\bi}\am_{\bi}$
are the spin and bath Hamiltonians, $F_{\bi}(\s)$ the spin-dependent
part of the interaction, and $V_{\bi}$ coupling constants.
For systems with discrete and finite spectrum, it is very convenient to
introduce the Hubbard (level-shift) operators
$X_{\n}^{\m}\equiv\nket\mbra$.
Any operator $F$ can then be expanded in this basis
$F
=
\sum_{\n\m}F_{\n\m}X_{\n}^{\m}$,
with $F_{\n\m}=\nbra F\mket$.
When coupled to the environment, the spin is not in a pure state and
it needs to be described by its density matrix $\dm$.
Its matrix elements are given in this framework by
$\dm_{\m\n}=\llangle X_{\n}^{\m}\rrangle$.


Many problems in quantum optics, magnetism, or chemical physics
involve {\em weak\/} system-bath coupling \cite{weiss}.
Then, the dynamical equation for $\dm$ can be obtained by perturbation
theory.
In the Hubbard formalism, using the $\Sz$ eigenstates as basis,
$\Sz\mket=\m\mket$, one finds for
$\Hs=\Hs_{\rm d}(\Sz)-\B\cdot\s$
the density-matrix equation
\cite{gar91llb,garchu97}
%
\begin{eqnarray}
\label{DME:nonmarkov}
\frac{\D{}}{\D t}
X_{\n}^{\m}
&=&
\iu\,
\tf_{\n\m}
X_{\n}^{\m}
+
(\iu/2)
\Bcp
\left(
\lf_{\m}
\,
X_{\n}^{\m+1}
-
\lf_{\n-1}
\,
X_{\n-1}^{\m}
\right)
\nonumber\\
&+&
(\iu/2)
\Bcm
\left(
\lf_{\m-1}
\,
X_{\n}^{\m-1}
-
\lf_{\n}
\,
X_{\n+1}^{\m}
\right)
+
R_{\n}^{\m}
\;.
\end{eqnarray}
The $\tf_{\n\m}\equiv\el_{\n}-\el_{\m}$ are the frequencies associated
with the $\m\to\n$ transition, being $\el_{\m}$ the levels of the
diagonal part of the spin Hamiltonian (including $\Bz$).
%
%
The circular components of the transverse field are
$\Bcpm=\Bx\pm\iu\By$ and the $\lf_{\m}=[S(S+1)-\m(\m+1)]^{1/2}$ are
ladder factors.

The first three terms in Eq.~(\ref{DME:nonmarkov}) give the {\em
unitary\/} evolution of the isolated spin in the Heisenberg
representation.
The {\em relaxation\/} term $R_{\n}^{\m}$ incorporates the effects of
the bath and has a non-Markovian (history dependent) form
%
\begin{eqnarray}
\label{Rnonmarkov}
R_{\n}^{\m}
=
-
\int_{-\infty}^{t}
\!\!\D{\tp}
&\big\{&
\Ker(\tp-t)
\,
F(\tp)
\,
\big[
F
\,,
X_{\n}^{\m}
\big]
\nonumber\\
& &
-
\Ker(t-\tp)
\,
\big[
F
\,,
X_{\n}^{\m}
\big]
\,
F(\tp)
\;
\big\}
\;.
\end{eqnarray}
Here the operators without time argument are evaluated at $t$ whereas
$F(\tp)=\sum_{\n'\m'}F_{\n'\m'}X_{\n'}^{\m'}(\tp)$.
The memory kernel $\Ker$ is given in terms of the {\em spectral
density\/} of bath modes,
$J(\omega)
\equiv
\tfrac{\pi}{2}
\sum_{\bi}
|V_{\bi}|^{2}
\,
\delta(\omega-\omega_{\bi})$,
and bosonic occupation numbers,
$n_{\omega}=(\e^{\omega/\kT}-1)^{-1}$,
by
%
\begin{equation}
\label{kernel:J}
\Ker(\tp)
=
\int_{0}^{\infty}
\!
\frac{\D\omega}{\pi}
\,
J(\omega)
\big[
n_{\omega}
\,
\e^{+\iu\omega\tp}
+
(n_{\omega}+1)
\,
\e^{-\iu\omega\tp}
\big]
\;.
\end{equation}

To second order in the interaction, and not too strong transverse
field, the {\em retarded\/} time dependences $X_{\n'}^{\m'}(\tp)$ can be
determined by the dominant term in the conservative evolution
$X_{\n}^{\m}(\tp)
\simeq
\e^{-\iu\tf_{\n\m}(t-\tp)}\,X_{\n}^{\m}(t)$.
%
%
Inserting such $X_{\n'}^{\m'}(\tp)$ in $R_{\n}^{\m}$, only operators
evaluated at $t$ remain and non-Markovian features effectively
disappear.
Then, the coefficients of the $X_{\n'}^{\m'}$ include, along with the
coupling matrix elements $F_{\n\m}$, the relaxation rates
$\Wu_{\n\tm\m}\equiv\Wu(\tf_{\n\m})$,
with the universal {\em rate function\/} associated to the kernel
$\Wu(\tf)
=
\mathrm{Re}
[\int_{0}^{\infty}\!\D{\tp}\,\e^{-\iu\tf\tp}\,\Ker(\tp)]$.
%

We shall consider in the sequel the following family of couplings
$F$: linear in $\Scpm=\Sx\pm\iu\Sy$ but allowing for
$\Sz$-dependent ``coefficients'' $\Vz(\Sz)$:
%
\begin{equation}
\label{F:lin}
F(\s)
=
\eta_{+}
[\Vz(\Sz),\,\Scm]_{+}
+
\eta_{-}
[\Vz(\Sz),\,\Scp]_{+}
\;.
\end{equation}
Here $\eta_{\pm}$ are some scalars ensuring $F^{\adj}=F$ while
$[A,\,B]_{+}\equiv A\,B+B\,A$.
Then, the matrix elements $F_{\n\m}=\nbra F\mket$ read
%
$F_{\n\m}
=
\LF_{\m,\m-1}
\,
\delta_{\n,\m-1}
+
\LF_{\m+1,\m}^{\ast}
\,
\delta_{\n,\m+1}$,
where
$\LF_{\m,\m'}
=
\eta_{+}[\Vz(\m)+\Vz(\m')]\lf_{\m,\m'}$
and $\lf_{\m,\m\pm1}=[S(S+1)-\m(\m\pm1)]^{1/2}$.
%
%
The relaxation term for these couplings acquires the following
(Redfield) form \cite{gar91llb,garchu97,garzue2005}
%
\begin{eqnarray}
\label{Rmarkov:lin:invk}
R_{\n}^{\m}
&=&
\LF_{\n,\n-1}\LF_{\m,\m-1}^{\ast}
\;
(\Wu_{\n\tm\n-1}+\Wu_{\m\tm\m-1})
\;
X_{\n-1}^{\m-1}
\nonumber\\
&-&
\big(\;\;\;
|\LF_{\n+1,\n}|^{2}
\;
\Wu_{\n+1\tm\n}
+
|\LF_{\m+1,\m}|^{2}
\;
\Wu_{\m+1\tm\m}
\nonumber\\
& &
\;\;
+
|\LF_{\n,\n-1}|^{2}
\;
\Wu_{\n-1\tm\n}
+
|\LF_{\m,\m-1}|^{2}
\;
\Wu_{\m-1\tm\m}
\;\;
\big)
\;
X_{\n}^{\m}
\nonumber\\
&+&
\LF_{\n+1,\n}^{\ast}\LF_{\m+1,\m}
\;
(\Wu_{\n\tm\n+1}+\Wu_{\m\tm\m+1})
\;
X_{\n+1}^{\m+1}
\;.
\end{eqnarray}
Inserting this $R_{\n}^{\m}$ in Eq.~(\ref{DME:nonmarkov}) we get the
master equation for our problem within a fully quantum treatment (no
phenomenological relaxation is introduced, no preconceived form of the
equation is assumed).
Recall finally that handling the spin precession requires to solve the
full density-matrix equation, because it involves off-diagonal
elements, and it is not captured by a (Pauli) master equation for the
level populations ($\dm_{\m\m}$).


We mentioned the difficulties to solve our models for quantum
dissipation and that the continued-fraction method (a relative of the
recursion method and Lanczos tri-diagonalization) has been applied to
several quantum systems \cite{allaribam77,shiuch93,vogris88,gar2004}.
For spin problems \cite{garzue2005} one starts writing the master
equation compactly as
$\dot{X}_{\n}^{\m}
=
\sum_{\n'\m'}\Q_{\n,\n'}^{\m,\m'}\,X_{\n'}^{\m'}$
with
$\n'=\n-1,~\n,~\n+1$
and 
$\m'=\m-1,~\m,~\m+1$.
To convert this 2-index differential recurrence into a 1-index one, we
introduce appropriate $(2S+1)$-vectors, $\mc_{\n}$, and
$(2S+1)\times(2S+1)$-matrices, $\mQ_{\n,\n'}$, with components and
elements \cite{garzue2005}
%
\begin{equation}
\label{vectors:matrices}
\big(\mc_{\n}\big)_{\m}
=
\llangle
X_{\n}^{\m}
\rrangle
\qquad
\big(\mQ_{\n,\n'}\big)_{\m\m'}
=
\Q_{\n,\n'}^{\m,\m'}
\;,
\end{equation}
obtaining
$\dot{\mc}_{\n}
=
\mQ_{\n,\n-1}\,\mc_{\n-1}+\mQ_{\n,n}\,\mc_{\n}+\mQ_{\n,\n+1}\,\mc_{\n+1}$.
In this 1-index form the recurrence can be tackled by (matrix)
continued-fraction methods \cite{risken} yielding the solution of the
master equation~(\ref{DME:nonmarkov}).
We have then the full density matrix
$\dm_{\m\n}=\langle X_{\n}^{\m}\rangle=(\mc_{\n})_{\m}$
and {\em any\/} observable (magnetization, susceptibilities, etc.) can
be computed from the trace formula
$\langle A\rangle=\Tr(\dm\,A)$.
The matrix associated to the original system, $\dot{\X}=\Q\,\X$, had
dimensions $(2S+1)^{2}\times(2S+1)^{2}$ difficulting the handling even
of moderate spins ($S\lesssim10$).
The continued-fraction approach replaces it by $2S+1$ problems with
matrices $(2S+1)\times(2S+1)$, allowing to gain some orders of
magnitude in $S$ and pursue the classical limit a way longer.


We now apply the above formalism to the problem of spin dynamics in a
magnetic-anisotropy plus Zeeman potential
$\Hs=-D\,\Sz^{2}-\B\cdot\s$.
This Hamiltonian may also be viewed as the minimal model for
superparamagnets \cite{blupra2004}.
%
The anisotropy term has two minima at $\Sz=\pm S$ with a barrier at
$\Sz=0$.
The coupling to the environment provokes quantum Brownian motion of
the spin, which may overcome the potential barriers.
We consider two basic solid-state mechanisms
\cite{weiss,legetal87}:
(i) Coupling to electron-hole excitations near the Fermi surface (a
bosonizable bath); then $F(\s)=\half(\eta_{+}\Scm+\eta_{-}\Scp)$
[i.e., $\Vz(\Sz)=\textrm{const}$ in Eq.~(\ref{F:lin})] while the bath
is {\em Ohmic}, $J(\omega)=\Wo\,\omega$.
(ii) Coupling to phonons; now $\Vz(\Sz)\propto\Sz$ and the environment
is {\em super-Ohmic}, $J(\omega)=\Wo\,\omega^{3}$ (in 3D).
The rates $\Wu_{\n'\tm\m'}$ required in $R_{\n}^{\m}$ can be obtained
from the spectral density by
$\Wu(\tf)=J(\tf)\,n_{\tf}+J(-\tf)(n_{-\tf}+1)$,
understanding $J(\omega<0)\equiv0$.
We start with the super-Ohmic nonlinear case, which has received less
attention in the context of quantum dissipative systems than the Ohmic
bilinear coupling; we will see that it also has a rich physics.

\begin{figure}[!]
\resizebox{8.cm}{!}{\includegraphics{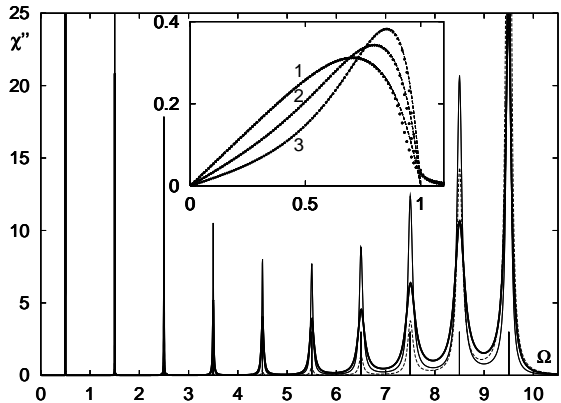}}
\caption{
Absorption line-shape $\chi''(\w)$ for a spin $S=10$ with $D=0.5$ at
$\Bz=0$.
Thick line: $\sigma=D\,S^{2}/\kT=5$ and spin-{\em phonon\/} coupling
$\Wo=3\cdot10^{-8}$.
Thin lines: effects of halving the damping at the same $T$ (solid),
and halving $T$ keeping $\Wo$ (dashed).
Vertical lines: loci of the transition frequencies
$\tf_{\m,\m+1}=D(2\m+1)$.
Inset: {\em classical\/} dampingless asymptote (\ref{chi:GG}) for
$\sigma=1$, $2$ and $3$ (lines), and exact Fokker--Planck results for
{\em finite\/} Landau-Lifshitz damping ($\lambda_{\textrm{LL}}=0.003$;
symbols).
}
\label{fig:SR:didactics}
\end{figure}

The Zeeman term and $F\propto\Scpm$ have non-zero matrix elements
between the states $\mket$, producing transitions between them.
In an oscillating field, they result in peaks in the imaginary part
$\chi''(\w)$ of the dynamical susceptibility (absorption line-shape)
located at the transition frequencies
$\tf_{\m,\m+1}=\el_{\m}-\el_{\m+1}=D(2\m+1)+\Bz$
(Fig.~\ref{fig:SR:didactics}).
The transitions at the potential wells ($|\m|\sim S$) correspond to
the largest frequencies ($\tf\sim2DS$ at $\Bz=0$), while those near
the barrier top ($\m\sim0$) appear at low $\w$ ($\sim D$).
Then, going from high to low $\w$, the intensity of the peaks
decreases, as they involve transitions between higher levels, which
are thermally less populated.

The peaks have finite width and height due to the damping $\Wo$ and
the temperature, as the interaction with the bath ``blurs'' the spin
energy levels.
Thus, a lowering of $\Wo$ or $T$ makes the peaks narrower and higher
(phenomenology akin to that of a damped oscillator).
There is an extra narrowing of the low $\w$ peaks, because the
spin-phonon coupling $F\sim\Sz\Scpm$ leads to an effective damping
decreasing with $\m$ [$\Wo_{\rm eff}\sim\Wo(2\m\pm1)^{2}$].
This enters in $R_{\n}^{\m}$ via the modified ladder factors
$|\LF_{\m,\m\pm1}|^{2}\sim(2\m\pm1)^{2}\lf_{\m,\m\pm1}^{2}$ and it is
the spin analogue of position-dependent damping in translational
Brownian motion.


Next, let us briefly discuss the corresponding {\em classical\/}
behavior.
The actual line-shape will depend on the phenomenological relaxation
model considered (Bloch equations, Landau--Lifshitz, etc).
Nevertheless, the result in the limit of zero damping is universal
\cite{gekh83e,garishpan90e}
%
\begin{equation}
\label{chi:GG}
\chi''(\w)
=
\frac{\mu^{2}}{\kT}
\frac{\pi}{2\Z}
\;
\w
\,
[1-(\w/\wa)^{2}]
\,
\exp[\sigma(\w/\wa)^{2}]
\;.
\end{equation}
Here $\Z$ is the partition function, $\wa$ the resonance frequency at
the wells, and $\sigma$ the barrier over $T$.
Physically, the anisotropy $\Hs_{\rm d}=-D\,\Sz^{2}$ leads to
$\Sz$-dependent precession frequencies
and the ensuing {\em spreading\/} of
the spectral line-shape (inset of Fig.~\ref{fig:SR:didactics}).
The population of the different $\Sz$-orbits changes with $T$,
modifying $\chi''(\w)$.
%
%
Note that this dissipationless limit provides a good description for
weak enough coupling in most of the $\w$ range.

We thus see that the classical phenomenology looks quite different
from the multi-peaked structure of the quantum case.
This poses the following questions: (i)~How does quantum mechanics
manage to join those two behaviors? and (ii)~which are the main
factors determining the way in which the classical phenomenology
emerges?
We now try to answer these questions by solving the density-matrix
equation~(\ref{DME:nonmarkov}) for increasing $S$ and getting as close
as possible to the classical domain.
\begin{figure}[b]
\resizebox{8.cm}{!}{\includegraphics{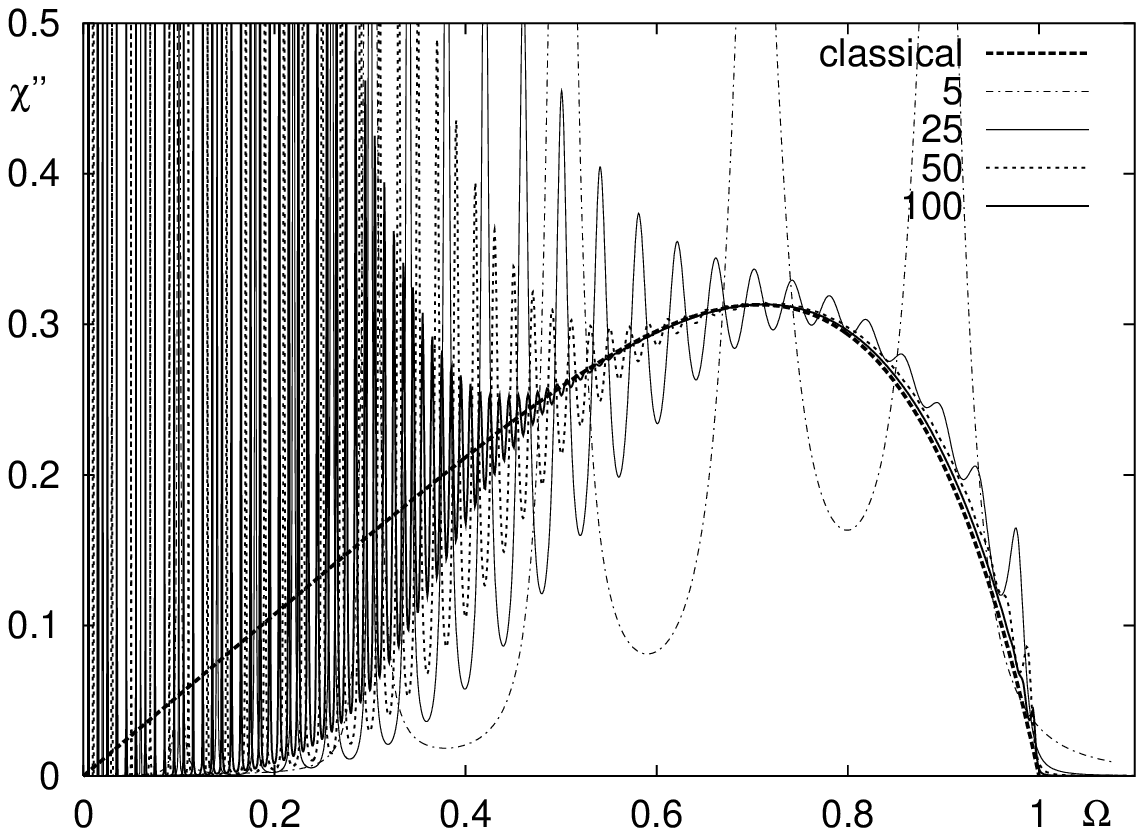}}
\resizebox{8.cm}{!}{\includegraphics{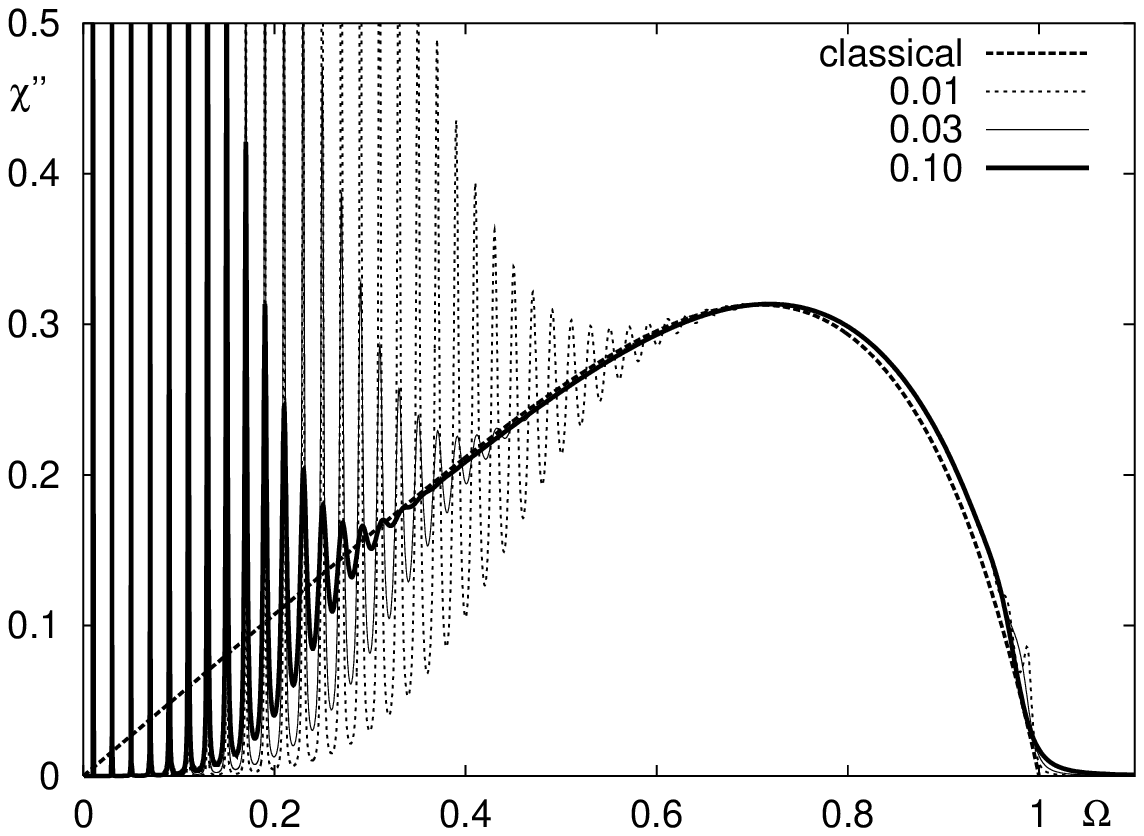}}
\caption{
Spectral line-shape $\chi''(\w)$ for $\sigma=1$ at $\Bz=0$
(see the text for the scaled quantities).
%
The thick dashed line is the classical Eq.~(\ref{chi:GG}).
Top: $S=5$, $25$, $50$, and $100$ with ``constant'' coupling
$\Wo/S=10^{-2}$.
Bottom: fixed $S=50$ with $\Wo/S=10^{-2}$ (as in top),
$3\cdot10^{-2}$, and $10^{-1}$.
}
\label{fig:SR:Q-cl}
\end{figure}


Recall, however, that limiting procedures in physical problems (e.g.,
lattice to continuous limit in field theories, thermodynamic limit in
statistical mechanics, etc.) require to define clearly:
(i) which quantities are kept constant when taking the limit and (ii)
which scaled variables are needed to monitor the evolution.
We fix the reduced anisotropy and field parameters
$\sigma=D\,S^{2}/\kT$
and
$\xi=S\,B/\kT$.
At constant $T$ this entails keeping the anisotropy-barrier height and
amount of Zeeman energy constant (and hence finite) while introducing
more levels with $S$ (the spacing decreases as $\tf\sim1/S$).
As for the scaled quantities, guided by the classical result
(\ref{chi:GG}), we use $\chi/\chi_{0}$ with $\chi_{0}=S(S+1)/\kT$
(corresponding to $\mu^{2}/\kT$) and $\w/2DS$ (which tends to
$\w/\wa$).
Finally, we also scale the bare coupling strength $\Wo$ with $S$.
Inspecting the density-matrix equation we see that the Hamiltonian
coefficients go as $\tf\sim1/S$ while the relaxation ones decrease as
$\Wo D^{2}\LF^{2}\tf^{2}\sim\Wo/S^{2}$ (we include a $D^{2}$
dependence arising in the coupling to phonons \cite{garchu97}).
Thus, fixing $\Wo/S$ we can study the effects of going to large $S$
while maintaining the relative ``weights'' of the conservative and
relaxation terms in the quantum master equation.

Proceeding in this way, we compute the transverse dynamical response
for various $S$ (Fig.~\ref{fig:SR:Q-cl}).
For moderate spins we clearly recognize the quantum features of
Fig.~\ref{fig:SR:didactics}.
As $S$ is increased more peaks are introduced into the same interval
$\w/\wa$.
Due to their finite width they start to coalesce and a limit curve
progressively emerges.
However, the approach is far from uniform in $\w$.
At low frequencies the peaks merge slowly with $S$; they are sharp and
narrow due to the $\m$-dependent damping associated to $\Sz$ in
$F\sim\Sz\Scpm$.
This is less relevant at high frequencies [large $|\m|$, $\delta
\Wo_{\rm eff}/\Wo_{\rm eff}\sim4/(2\m\pm1)$] and a smooth peakless
line-shape arises there.
For a fixed $S$, in addition, one would expect that larger spin-bath
coupling will ``accelerate'' the classical convergence.
Figure~\ref{fig:SR:Q-cl} actually shows that the wildly peaked part is
then pushed further into the low $\w$ sector and that the
``oscillations'' around the limit curve are reduced.
It is remarkable that this limit curve is indeed Gekht's classical
prediction~(\ref{chi:GG}).
%

Finite width of the absorption peaks has been essential to reconstruct
the classical curve.
Here it has been provided by the coupling to the environment; in other
situations different broadening mechanisms may contribute
\cite{white}.
The form of the interaction, on the other hand, has led to a highly
non-uniform approach to the classical asymptote.
This shows that not only the strength, but also the structure of the
coupling Hamiltonian can play an important role in the approach to the
classical regime.
\begin{figure}[!]
\resizebox{8.cm}{!}{\includegraphics{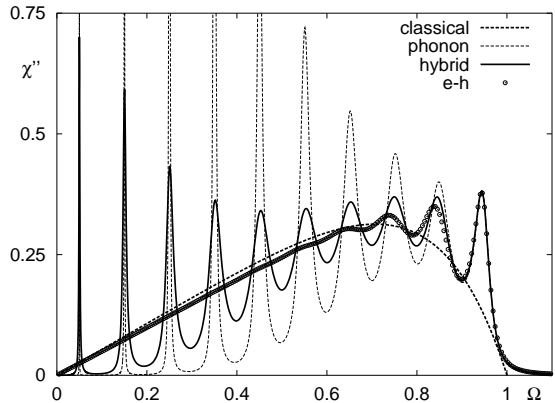}}
\caption{
Line-shape for $S=10$ at $\sigma=1$ with (i) the phonon-coupling model
and (ii) bilinear coupling, both with super-Ohmic (hybrid) and Ohmic
(electron-hole) spectral densities.
}
\label{fig:SR:Q-cl-ph-eh}
\end{figure}

This can be further supported by comparing with the ``electron-hole''
coupling model, where $F\sim\Scpm$.
To assess the different contributions we proceed in two steps,
adjusting $\Wo$ to get the same intensity for the ground-state
transitions; Fig.~\ref{fig:SR:Q-cl-ph-eh}.
First, we go from the phonon-coupling $F\sim\Sz\Scpm$ to a hybrid
model with $F\sim\Scpm$, but still super-Ohmic spectral density.
This $F$ greatly tames the low-frequency sharp peaks, but some
non-uniformity still remains, due to $\Wu_{\m\tm\m+1}\sim
J(\tf)\,n_{\tf}\sim\tf_{\m,\m+1}^{3-1}$ in the relaxation term.
Second, we add the Ohmic bath $J\propto\omega$ to the bilinear
coupling.
Then $\Wu_{\m\tm\m+1}\sim\tf_{\m,\m+1}^{1-1}\sim\textrm{const}$ and
the approach to the classical behavior becomes quite uniform in
most of the $\w$ range, in spite of the moderate spin value considered
($S=10$).

In summary, we have addressed the problem of spin dynamics in a
dissipative thermal bath.
Solving exactly the quantum master equation by a continued-fraction
method for increasing $S$ has allowed us to approach the classical
prediction for the absorption spectra.
We have investigated the effects of the spin-bath interaction on the
quantum-to-classical crossover.
The coupling strength, as usual in quantum dissipative systems,
accentuates the attainment of the classical phenomenology.
However, the approach is qualitatively affected by the structure of
the interaction, as illustrated with the different convergences in the
different sectors of the absorption spectra for two important
solid-state mechanisms.
Although the relevance of dissipation, specially in mesoscopic
systems, is amply recognized, only studies of decoherence and approach
to equilibrium had paid due attention to the structure of the coupling
Hamiltonian.
Here we have shown its relevance also in the features of the
quantum-classical border and in the emergence of classical behavior.


This work was partially supported by project BFM2002-00113 (DGES,
Spain).



\end{document}